\documentclass[%
reprint,
amsmath,amssymb, 
aps,
prb
]{revtex4-2}
\usepackage[utf8]{inputenc}
\usepackage{graphicx, color}
\usepackage{dcolumn}
\usepackage{bm}
\usepackage[colorlinks,urlcolor=blue,linkcolor=blue,anchorcolor=blue,citecolor=blue]{hyperref} 
\usepackage{ulem}
\usepackage{xcolor}

\begin{document}
\preprint{APS/123-QED}
\title{
 Unconventional compensated magnetic material LaMn$_2$SbO$_6$}
\author{Xiao-Yao Hou$^{1,2}$}
\author{Ze-Feng Gao$^{1,2}$}
\author{Huan-Cheng Yang$^{1,2}$}
\email{hcyang@ruc.edu.cn}
\author{Peng-Jie Guo$^{1,2}$}
\email{guopengjie@ruc.edu.cn}
\author{Zhong-Yi Lu$^{1,2,3}$}
\email{zlu@ruc.edu.cn}
\affiliation{1. School of Physics and Beijing Key Laboratory of Opto-electronic Functional Materials $\&$ Micro-nano Devices. Renmin University of China, Beijing 100872, China}
\affiliation{2. Key Laboratory of Quantum State Construction and Manipulation (Ministry of Education), Renmin University of China, Beijing 100872, China}
\affiliation{3. Hefei National Laboratory, Hefei 230088, China}

\date{\today}
\begin{abstract}
Unconventional magnetism including altermagnetism and unconventional compensated magnetism, characterized by its duality of real-space antiferromagnetic alignment and momentum-space spin splitting, has garnered widespread attention. While altermagnetism has been extensively studied, research on unconventional compensated magnetism remains very rare. In particular, unconventional compensated magnetic materials are only theoretically predicted and have not yet been synthesized experimentally. In this study, based on symmetry analysis and the first-principles electronic structure calculations, we predict that LaMn$_2$SbO$_6$ is a unconventional compensated magnetic semiconductor. Given that the Mn ions at opposite spin lattice cannot be connected by any symmetry, the spin splitting in LaMn$_2$SbO$_6$ is isotropic. More importantly, LaMn$_2$SbO$_6$ has already been synthesized experimentally, and its magnetic structure has been confirmed by neutron scattering experiments. Therefore, LaMn$_2$SbO$_6$ serves as an excellent material platform for investigating the novel physical properties of unconventional compensated magnetic materials.
\end{abstract}

\maketitle

{\it Introduction.} 
Conventionally, magnetism is usually divided into two major categories: ferromagnetism and antiferromagnetism. Ferromagnetic materials have non-relativistic spin-polarized splitting electronic band structures due to breaking time-reversal symmetry ($T$) and have been used widely in spintronic devices. However, the devices based on ferromagnetic materials experience several shortcomings: (i) they are easily affected by the external magnetic field and temperature, (ii) the information storage density is low, and (iii) the device speed is relatively slow (GHz magnitude). In contrast, the information storage density of antiferromagnetic materials can reach the atomic level, and the device speed based on antiferromagnetic materials can be up to THz range. However, the spin polarization of antiferromagnetism is negligible because of its zero net magnetic moment, which means that antiferromagnetism is inefficient in generating and transmitting spin polarized currents, which is not conducive to the performance improvement of spintronic devices. The desirable materials for next generation spintronic devices should take the advantages of both ferromagentic and antiferromanetic materials. 


Recently, altermagnetism, as a newly discovered third type of magnetism, has attracted a great deal of attention in the field of magnetism\cite{altermagnetism-1,altermagnetism-2,altermagnetism-3,altermagnetism-4}. In altermagnetic materials, the opposite spin sublattices cannot be connected by space-inversion symmetry and fractional translation symmetry, but by rotation or mirror symmetry. Altermagnet has time-reversal symmetry-breaking responses and spin polarization phenomena, 
such as the giant magnetoresistance (GMR) effect\cite{GMR-2024,GMR-PRX2022}, the anomalous Hall effect (AHE)\cite{AHE-hou2023}, the quantum anomalous Hall effect (QHE)\cite{QAH-npj2023}, and so on. Moreover, several altermagnetic materials have been confirmed experimentally\cite{exp-1,exp-3}. 


Although altermagnetism has the advantages of both ferromagnetism and antiferromagnetism, it has to be admitted that because the crystal field anisotropy of magnetic atoms with opposite spins is weak in the majority of altermagnetic materials, the spin splitting is correspondingly small\cite{tan2024}, which brings difficulties to practical applications. This motivates the search for unconventional compensated magnetic materials. In unconventional compensated magnetism, two opposite spin sublattices cannot be connected by any symmetry operation, resulting in isotropic spin splitting just like ferromagnetism, but the net magnetic moment is still zero just like antiferromagnetism according to Luttinger's theorem\cite{Luttinger1,Luttinger2,Luttinger3}. Although several unconventional compensated magnetic materials have been theoretically predicted, none of these materials have been synthesized experimentally\cite{Cheng-Cheng-PRL,Guo2025Luttinger,PhysRevLett.133.216701,PhysRevLett.132.156502,PhysRevLett.74.1171,PhysRevX.12.040002}.


In this study, based on symmetry analysis and the first-principles electronic band structure calculations , we predicted LaMn$_2$SbO$_6$ is a unconventional compensated magnetic semiconductor. 
Due to the absence of any symmetry connecting the Mn atoms with opposite magnetic moments, LaMn$_2$SbO$_6$ exhibits isotropic $s$-wave spin splitting. Moreover, under spin-orbit coupling (SOC), the top of valence bands of LaMn$_2$SbO$_6$ still largely maintain complete spin polarization. The N\'{e}el temperature of LaMn$_2$SbO$_6$ is estimated to be around 50 K by Monte Carlo simulation. 



\textit{Computational method.} 
The structural optimization and electronic structure calculation of LaMn$_2$SbO$_6$ were studied in the framework of density functional theory (DFT) \cite{37, 38, 39} by using the Vienna Ab initio Simulation Package (VASP) \cite{40, 41, 42}. The core electrons as well as the interaction between the core and valence electrons were described by using the projector augmented-wave method \cite{43}. The generalized gradient approximation (GGA) of Perdew-Burke-Ernzerhof (PBE) \cite{44} type was adopted as the exchange-correlation functional, using an energy cut-off of 600 eV for the plane waves. A 6 × 6 × 6 Monkhorst-Pack k mesh was used for the Brillouin zone (BZ) sampling of the unit cell. The internal atomic positions were fully relaxed until the forces on all atoms were smaller than 0.01 eV/Å. In order to take into account the correlation effects of Mn 3d orbitals, we carried out GGA+U \cite{45} calculations employing the simplified rotationally invariant version proposed by Dudarev et al. \cite{46}. The onsite effective $U_{eff}$ value of Mn 3d electrons in LaMn$_2$SbO$_6$ was choose as 3.8 eV. The Monte Carlo simulations based on the classical Heisenberg model were conducted by making use of the open source project MCSOLVER \cite{47}. 

 \begin{figure}[htbp]
	\centering
	\includegraphics[width=9 cm]{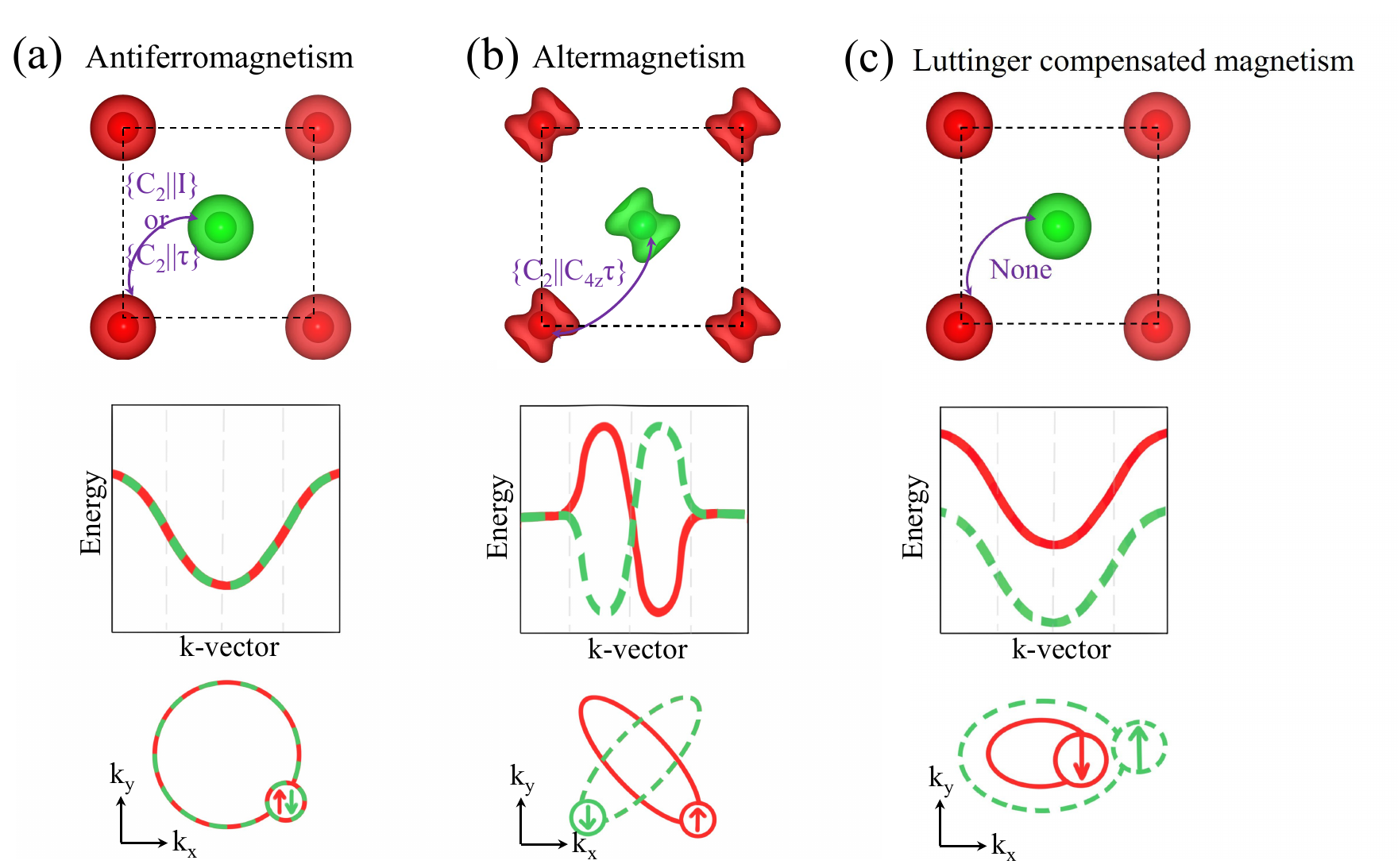}
	\caption{Schematic diagrams of the three types of compensated magnetism. (a) antiferromagnetism, (b) altermagnetism, (c) unconventional compensated magnetism. The upper, middle, and lower panels in (a), (b), and (c) represent the polarized charge density, band structure, and Fermi surface, respectively. The red and green represent spin-up and spin-down channels, respectively.}
	\label{Fig. 5}
\end{figure}
\textit{Results.} Based on spin group symmetry, collinear compensated magnetism can be classified into three categories as shown in Fig. 1. The first category is conventional collinear antiferromagnetism, whose sublattices with opposite spin moments are connected by $\left\{C_2||I\right\}$ or $\left\{C_2||\tau\right\}$ spin symmetry (Fig. 1(a)), where $I$ and $\tau$ represent space-inversion and fractional translation operations, respectively. These symmetries enforce Kramers degeneracy throughout the entire BZ, thus conventional collinear antiferromagnetic materials have spin-degenerate electronic bands without spin-orbit coupling (SOC) as shown in Fig. 1(a). The second category is altermagnetism. In altermagnetism, the sublattices with opposite spin moments cannot be connected by $\left\{C_2||I\right\}$ and $\left\{C_2||\tau\right\}$ spin symmetry but by $\left\{C_2||R\right\}$ or mirror $\left\{C_2||M\right\}$ spin symmetries (Fig. 1(b)), which guarantees zero total magnetic moment. The breaking of $\left\{C_2||I\right\}$ and $\left\{C_2||\tau\right\}$ lifts Kramers degeneracy, leading to band structure with anisotropic spin splitting without SOC (Fig. 1(b)). The third category is the unconventional compensated magnetism, where the opposite spin sublattices are not connected by any symmetry (Fig. 1(c)). Therefore, the electronic band structure is isotropic $s$-wave spin splitting without SOC (Fig. 1(c)). Compared to altermagnetism, unconventional compensated magnetism has not yet been experimentally confirmed. Here, we aim to predict unconventional compensated magnetic materials among the materials that have already been synthesized.

LaMn$_2$SbO$_6$ has been reported as a doubly ordered perovskite structure (also reported as double-double perovskite,  DDPv) with space group $P4_2/n$ (No. 86). The elementary symmetry operations of the $P4_2/n$ space group are $C_{4z}(\frac{1}{2},  \frac{1}{2},  \frac{1}{2})$ and $I (\frac{1}{2},  \frac{1}{2},  \frac{1}{2})$, resulting in the point group $C_{4h}$. A schematic diagram of the unit cell is shown in Fig. 2(a), which combines the columnar (Mn and La atoms) and rock-salt (Mn and Sb atoms) orders of 1:1 cations, respectively. The corresponding BZ together with the high-symmetry points and high-symmetry lines are shown in Fig. 2(b).
\begin{figure*}[htbp]
	\centering
	\includegraphics[width=17 cm]{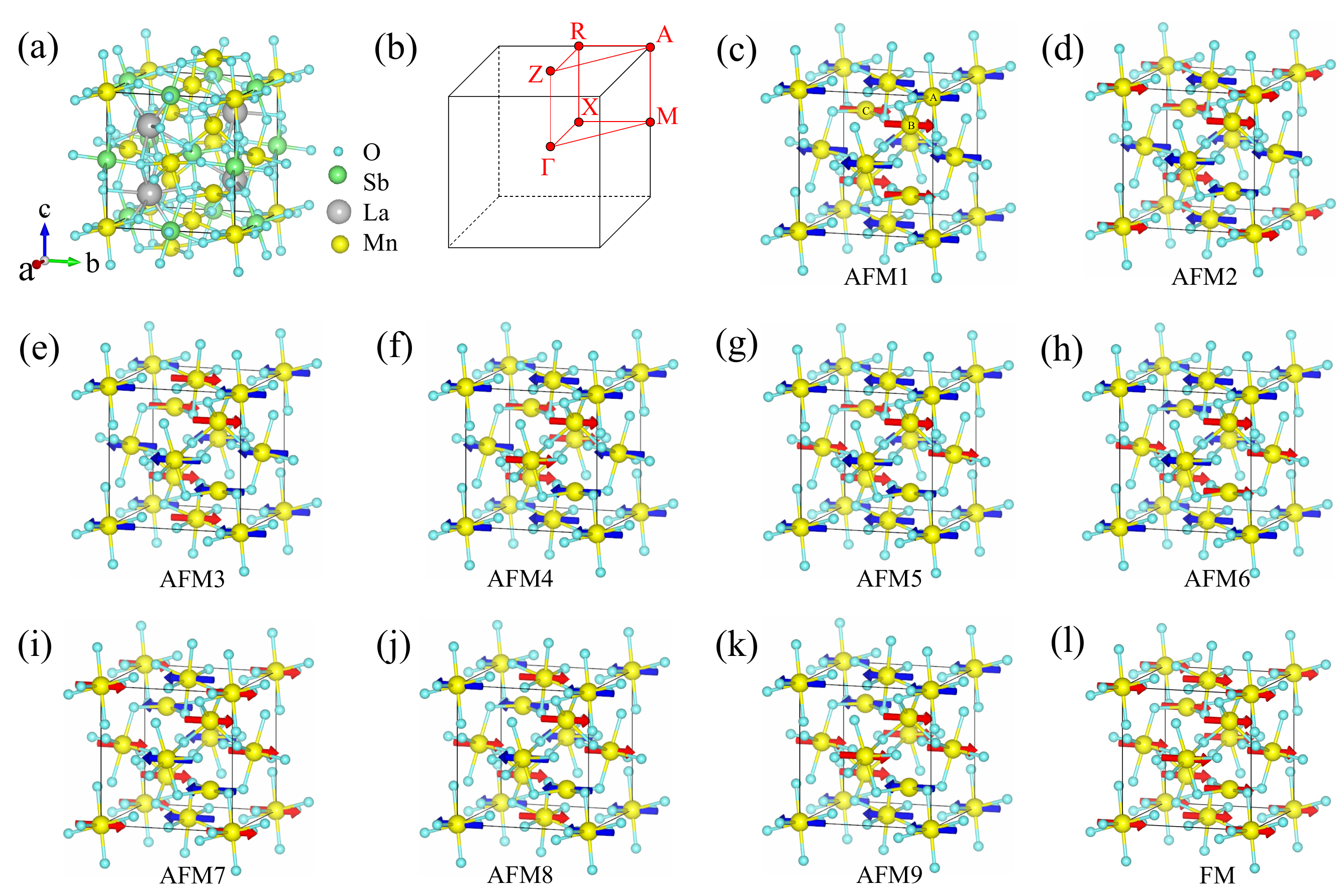}
	\caption{(a) Nonmagnetic crystal structure of LaMn$_2$SbO$_6$. (b) The corresponding bulk Brillouin zone,  the high-symmetry points and paths are marked by red dots and lines. (c-k) nine trial AFM configurations. (l) FM configuration. }
	\label{Fig. 1}
\end{figure*}

In LaMn$_2$SbO$_6$, the Mn atoms are respectively located in octahedral, tetrahedral, and square planar crystal fields, which are correspondingly labeled as A, B, and C as shown in Fig. 2(c). Due to the complexity of the atomic structure of LaMn$_2$SbO$_6$, 
we considered ten magnetic configurations to check the magnetic ground state of LaMn$_2$SbO$_6$ as shown in Fig. 2(c) - 2(l). 
Our calculations show that the total energy of the AFM1 magnetic state is always the lowest among the ten magnetic configurations with the variation of correlation interaction (Fig. 3(a)). Therefore, the AFM1 state is the magnetic ground state of LaMn$_2$SbO$_6$, which is consistent with the results of neutron scattering experiment \cite{MnLaMnSbO,MnLaMnSbO-2}. The AFM1 state is intra-layer ferromagnetic and inter-layer antiferromagnetic (Fig. 2(c)) and the magnetic primitive cell for AFM1 state is the same as the crystal primitive cell. So  LaMn$_2$SbO$_6$ has no $\left\{C_2||\tau\right\}$ symmetry. Furthermore, our symmetry analysis reveals that all symmetry operations in the nonsymmorphic C$_{4h}$ point group can only connect the Mn ions with the same spin moments for LaMn$_2$SbO$_6$. 
Thus, the Mn ions with opposite spin moments can not be connected by any symmetry. Therefore, LaMn$_2$SbO$_6$ may be a unconventional compensated magnetic material.

On the other hand, the magnetic transition temperature is a very important feature for magnetic materials. To estimate the Néel temperature of LaMn$_2$SbO$_6$, we begin by considering a classical Heisenberg model that includes the  first- and second-nearest-neighbor exchange interactions as a try. The exchange interaction parameters J$_1$ and J$_2$ can be obtained by mapping the energies of three distinct magnetic structures onto the tentative Heisenberg model. Specifically, we select the ground-state AFM1 configuration along with any two of AFM2, AFM3, AFM5, and AFM6. These four magnetic configurations closely resemble the ground state magnetic configuration AFM1. Our calculations reveal that the magnitude of J\textsubscript{1} is approximately two orders larger than that of 
J\textsubscript{2}. Thus, J$_2$ is negligible, we adopt a first-nearest-neighbour 3D cubic lattice classical Heisenberg model :
\begin{align}
H=-J\textsubscript{1}\sum_{<i, j>}S\textsubscript{i}S\textsubscript{j}
\end{align}
where S\textsubscript{i(j)} represents the spin of Mn\textsuperscript{2+} 
ion on site i(j). J\textsubscript{1} denote the first-nearest exchange interaction parameter between Mn(a) and Mn(b). 
Mapping the energies of the ground state magnetic configuration AFM1 and AFM6 onto the above Heisenberg model yields J\textsubscript{1}S$^2$. The values of J\textsubscript{1}S$^2$ for different U values are presented in Table 1. As illustrated in Figure 3(c), the Néel temperature of LaMn$_2$SbO$_6$ can be extracted from the peak of the specific-heat capacity based on the Monte Carlo simulations. When U=3.8 eV, the simulated Néel temperature is 48.99 K, which closely aligns with the experimental value ($\sim$ 48 K)\cite{MnLaMnSbO}.

\begin{table*}
    \caption{The values of J\textsubscript{1}S$^2$ at different U values.}
\begin{ruledtabular}
    \centering
    \begin{tabular}{ccccccccc}
         U (eV)& 0 & 1 & 2 & 3 & 4 & 5 & 6 & 7 \\
         \colrule
         J\textsubscript{1}S$^2$ (K) &-145.71& -110.91 & -87.00 & -69.99 & 57.46 & -47.43 & -39.89 & -33.63 \\
    \end{tabular}
    \label{tab:my_label}
    \end{ruledtabular}
\end{table*}

 \begin{figure}[htbp]
	\centering
	\includegraphics[width=9 cm]{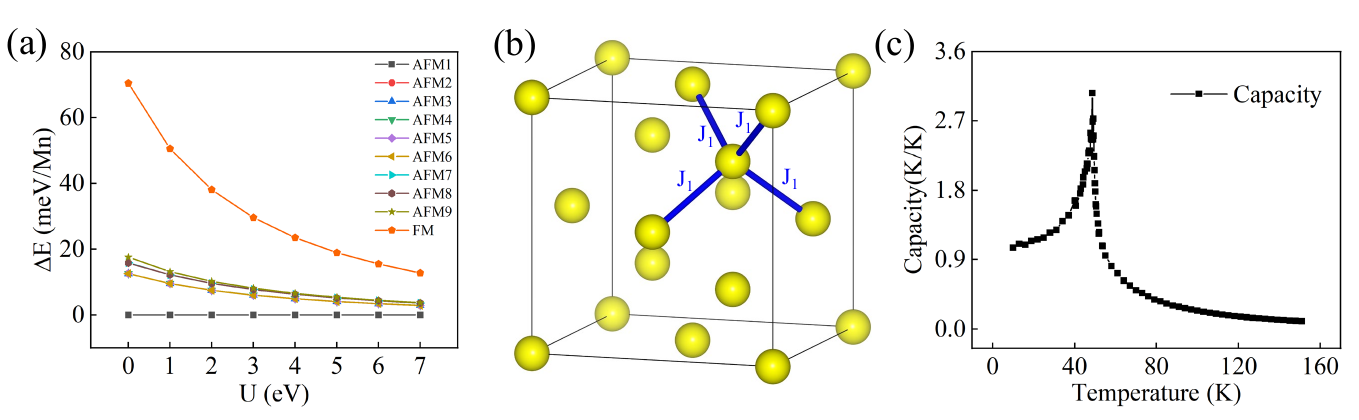}
	\caption{(a) Relative energies of eight trial AFM and the FM states with respect to the AFM1 state at different U values. (b) The blue lines represent the first- earest neighbors. (c) The evolution of specific heat capacity with temperature of LaMn$_2$SbO$_6$. }
	\label{Fig. 2}
\end{figure}

Next, we study the electronic properties of LaMn$_2$SbO$_6$ with the correlation interaction U set at 3.8 eV. Without SOC, LaMn$_2$SbO$_6$ is a direct bandgap semiconductor with bandgap being 1.57 eV, and its spin-up and spin-down bands are completely split. This is due to the absence of any symmetry connecting Mn ions with opposite magnetic moments. From Fig. 4(a), the valence bands of LaMn$_2$SbO$_6$ exhibit significant spin splitting, while the conduction bands show relatively smaller spin splitting. We speculate that the valence bands of LaMn$_2$SbO$_6$ are primarily contributed by the 3d orbitals of Mn atoms, while the conduction bands are mainly contributed by non-magnetic atoms. To confirm this, we plotted the band structure of LaMn$_2$SbO$_6$ with orbital weights as shown in Fig. 4(b). From Fig. 4(b) the valence bands of LaMn$_2$SbO$_6$ are indeed mainly contributed by the 3d orbitals of Mn, while the conduction bands are primarily contributed by Sb atoms. On the other hand, we also calculated the polarized charge density distribution of LaMn$_2$SbO$_6$. Although the Mn atoms are respectively in octahedral, tetrahedral, and square planar crystal fields of O atoms, the polarized charge densities of all Mn ions are approximately spherical, which reflects that all Mn are half filled and in high-spin state. The calculated magnetic moment on every Mn is about 4.6 $\mu$B, indicating strong Hund's interactions among Mn 3d electrons. 
Therefore, the orbital magnetic moments are quenched well in LaMn$_2$SbO$_6$. Additionally, since LaMn$_2$SbO$_6$ is a semiconductor with identical electrons for spin-up and spin-down, its total spin magnetic moment should be zero according to the Luttinger theorem\cite{Luttinger1,Luttinger2,Luttinger3}. Thus, LaMn$_2$SbO$_6$ is a unconventional compensated magnetic material with robust zero net magnetic moment.

 \begin{figure}[htbp]
	\centering
	\includegraphics[width=9 cm]{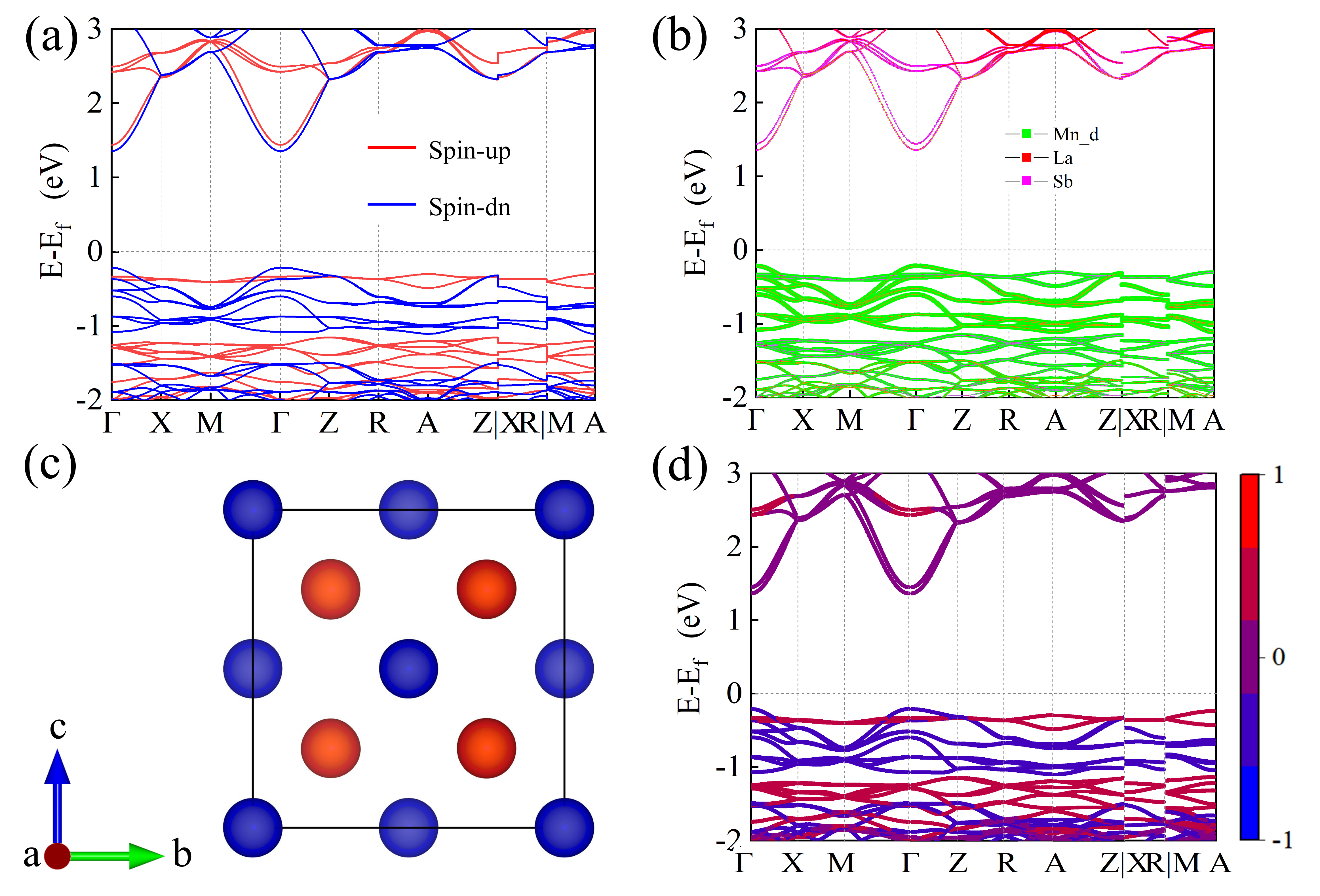}
	\caption{ (a) The electronic band structure without SOC at $U_{eff}$ = 3.8 eV. (b) The corresponding fat band without SOC. (c) The spin-polarized density for spin-up (red) and spin-dn (blue). (d) The electronic band structure with SOC at $U_{eff}$ = 3.8 eV.}
	\label{Fig. 3}
\end{figure}

When SOC is taken into account, the calculations of magnetic anisotropy indicate that the easy magnetization axis of LaMn$_2$SbO$_6$ is along the y-axis, which is consistent with experiments\cite{MnLaMnSbO}. Therefore, the spin expectation values of the bands are dominated by S$_y$, while S$_x$ and S$_z$ are small. Then, we have plotted the band structure with S$_y$ projections as shown in Fig. 4(d). Although SOC does not lead to significant changes for the bands, it does cause noticeable alterations for polarization of bands. From Fig. 4(d), the spin polarization of the bottom of conduction bands has changed significantly, which is attributed to the fact that these conduction bands are primarily contributed by the Sb atomic orbitals with strong SOC. In contrast, the valence bands, which are mainly contributed by the 3d orbitals of Mn atoms with weak SOC, still largely maintain complete spin polarization. In addition, our calculations also show that the magnetic moment of LaMn$_2$SbO$_6$ with spin-orbit coupling is negligible, only 0.001 Bohr magneton.

Although the bands at the $\Gamma$ point and the A point have opposite spin polarizations, the bands near the A point are lower than those at the $\Gamma$ point by 90 meV (Fig. 4(d)). Since doping in semiconductors is very low, only the spin-down holes at the $\Gamma$ point contribute to the polarized current. Therefore, LaMn$_2$SbO$_6$ will not only retain its semiconducting properties but also  generate a net spin current. Moreover, LaMn$_2$SbO$_6$ can avoid the disadvantage of stray fields caused by non-zero magnetic moments in ferromagnetic semiconductors. Therefore, LaMn$_2$SbO$_6$ may be very advantageous for the research of new electronic devices.

\textit{Summary.} Based on symmetry analysis and the first-principles electronic structure calculations, we have systematically investigated the magnetic and electronic properties of LaMn$_2$SbO$_6$. The calculated ground state magnetic structure is consistent with the results of neutron scattering experiments, and the electronic structure calculations show that LaMn$_2$SbO$_6$ is a unconventional compensated magnetic semiconductor with a 1.57 eV band gap. Due to the absence of any symmetry connecting Mn atoms with opposite magnetic moments, LaMn$_2$SbO$_6$ exhibits isotropic $s$-wave spin splitting. Moreover, under SOC, the valence bands of LaMn$_2$SbO$_6$ still largely maintain complete spin polarization. LaMn$_2$SbO$_6$ is an excellent material platform for studying the novel physical properties of unconventional compensated magnetism.

\begin{acknowledgments}
This work was financially supported by the National Natural Science Foundation of China
(Grant No.12204533, No.12434009 and No.62476278), the National Key R$\&$D Program of China (Grant No. 2024YFA1408601), the Fundamental Research Funds for the Central Universities, and the Research Funds of Renmin University of China (Grant No. 24XNKJ\textsubscript{1}5).Computational resources have been provided by the Physical Laboratory of High Performance Computing at Renmin University of China.
\end{acknowledgments}

\nocite{*}


\bibliographystyle{IEEEtran}

\begin{thebibliography}{99}\footnotesize
\itemsep=-1pt plus.2pt minus.2pt
\bibitem{altermagnetism-1} 
Hayami S, Yanagi Y and Kusunose H 2019 {\it J. Phys. Soc. Jpn.} {\bf 88} 123702
\bibitem{altermagnetism-2}
Šmejkal L, González-Hernández R, Jungwirth T and Sinova J 2020 {\it Sci. Adv.} {\bf 6} eaaz8809
\bibitem{altermagnetism-3}
Yuan L D, Wang Z, Luo J W, Rashba E I and Zunger A 2020 {\it Phys. Rev. B} {\bf 102} 014422
\bibitem{altermagnetism-4}
Mazin I I, Koepernik K, Johannes M D, González-Hernández R and Šmejkal L 2021 {\it Proc. Natl. Acad. Sci. U.S.A.} {\bf 118} e2108924118
\bibitem{GMR-2024}
Zhang R W, Cui C X, Li R Z, Duan J Y, Li L, Yu Z M and Yao Y G 2024 {\it Phys. Rev. Lett.} {\bf 133} 056401
\bibitem{GMR-PRX2022}
Šmejkal L, Hellenes A B, González-Hernández R, Sinova J and Jungwirth T 2022 {\it Phys. Rev. X} {\bf 12} 011028
\bibitem{AHE-hou2023}
Hou X Y, Yang H C, Liu Z X, Guo P J and Lu Z Y 2023 {\it Phys. Rev. B} {\bf 107} L161109
\bibitem{QAH-npj2023}
Guo P J, Liu Z X and Lu Z Y 2023 {\it npj Comput. Mater.} {\bf 9} 70
\bibitem{exp-1}
Krempaský J, Šmejkal L, D'Souza S W, Hajlaoui M, Springholz G, Uhlířová K, Al Arab F, Constantinou P C, Strokov V, Usanov D, Pudelko W R, González-Hernández R, Birkhellenes A, Jansa Z, Reichlová H, Šobán Z, Gonzalez Betancourt R D, Wadley P, Sinova J, Kriegner D, Minár J, Dil J H and Jungwirth T 2024 {\it Nature} {\bf626} 517
\bibitem{exp-3}
Reimers S, Odenbreit L, Šmejkal L, Strocov V N, Constantinou P, Hellenes A B, Ubiergo R J, Campos W H, Bharadwaj V K, Chakraborty A, Denneulin T, Shi W, Dunin-Borkowski R E, Das S, Kläui M, Sinova J and Jourdan M 2024 {\it Nat. Commun.} {\bf 15} 2116
\bibitem{tan2024}
Tan C Y, Gao Z F, Yang H C, Liu K, Guo P J and Lu Z Y 2024 arXiv:2406.16603
\bibitem{Luttinger1}
Luttinger J M and Ward J C 1960 {\it Phys. Rev.} {\bf 118} 1417
\bibitem{Luttinger2}
Kohn W and Luttinger J M 1960 {\it Phys. Rev.} {\bf 118} 41
\bibitem{Luttinger3}
Luttinger J M 1960 {\it Phys. Rev.} {\bf 119} 1153
\bibitem{Cheng-Cheng-PRL}
Liu Y, Guo S D, Li Y and Liu C C 2024 {\it Phys. Rev. Lett.} {\bf 134} 116703
\bibitem{Guo2025Luttinger}
Guo P J, Hou X Y, Gao Z F, Yang H C, Ji W and Lu Z Y 2025 arXiv:2502.18136
\bibitem{PhysRevLett.133.216701}
Yuan L D, Georgescu A B and Rondinelli J M 2024 {\it Phys. Rev. Lett.} {\bf 133} 216701
\bibitem{PhysRevLett.132.156502}
Kawamura T, Yoshimi K, Hashimoto K, Kobayashi A and Misawa T 2024 {\it Phys. Rev. Lett.} {\bf 132} 156502
\bibitem{PhysRevLett.74.1171}
van Leuken H and de Groot R A 1995 {\it Phys. Rev. Lett.} {\bf 74} 1171
\bibitem{PhysRevX.12.040002}
Mazin I 2022 {\it Phys. Rev. X} {\bf 12} 040002
\bibitem{37}
Kohn W and Sham L J 1965 {\it Phys. Rev.} {\bf 140} A1133
\bibitem{38}
Hohenberg P and Kohn W 1964 {\it Phys. Rev.} {\bf 136} B864
\bibitem{39}
Kresse G and Furthmüller J 1996 {\it Phys. Rev. B} {\bf 54} 11169
\bibitem{40}
Kresse G and Hafner J 1993 {\it Phys. Rev. B} {\bf 47} 558
\bibitem{41}
Kresse G and Furthmüller J 1996 {\it Comput. Mater. Sci.} {\bf 6} 15
\bibitem{42}
Kresse G and Joubert D 1999 {\it Phys. Rev. B} {\bf 59} 1758
\bibitem{43}
Perdew J P, Burke K and Ernzerhof M 1996 {\it Phys. Rev. Lett.} {\bf 77} 3865
\bibitem{44}
Marzari N, Mostofi A A, Yates J R, Souza I and Vanderbilt D 2012 {\it Rev. Mod. Phys.} {\bf 84} 1419
\bibitem{45}
Anisimov V I, Zaanen J and Andersen O K 1991 {\it Phys. Rev. B} {\bf 44} 943
\bibitem{46}
Dudarev S L, Botton G A, Savrasov S Y, Humphreys C J and Sutton A P 1998 {\it Phys. Rev. B} {\bf 57} 1505
\bibitem{47}
Mostofi A A, Yates J R, Pizzi G, Lee Y S, Souza I, Vanderbilt D and Marzari N 2014 {\it Comput. Phys. Commun.} {\bf 185} 2309
\bibitem{MnLaMnSbO}
Solana-Madruga E, Arévalo-López Á M, Dos Santos-García A J, Ritter C, Cascales C, Sáez-Puche R and Attfield J P 2018 {\it Phys. Rev. B} {\bf 97} 134408
\bibitem{MnLaMnSbO-2}
Solana-Madruga E, Arévalo-López Á M, Dos-Santos-García A J, Urones-Garrote E, Ávila-Brande D, Sáez-Puche R and Attfield J P 2016 {\it Angew. Chem. Int. Ed.} {\bf 55} 9340

\end{thebibliography}

\end{document}